\documentclass[%
 reprint,
 amsmath,amssymb,
 aps,
]{revtex4-2}

\usepackage{graphicx}
\usepackage{dcolumn}
\usepackage{bm}

\begin{document}

\preprint{APS/123-QED}

\title{Selective optical charge-state preparation of a quantum dot molecule\\ for independent control of orbital coupling}

\author{Frederik Bopp$^1$}
\author{Jonathan Rojas$^1$}
\author{Natalia Revenga$^1$}
\author{Hubert Riedl$^1$}
\author{Friedrich Sbresny$^2$}
\author{Katarina Boos$^2$}
\author{Tobias Simmet$^1$}
\author{Arash Ahmadi$^1$}
\author{David Gershoni$^3$}
\author{Jacek Kasprzak$^4$}
\author{Arne Ludwig$^5$}
\author{Stephan Reitzenstein$^6$}
\author{Andreas Wieck$^5$}
\author{Dirk Reuter$^7$}
\author{Kai Müller$^2$}
\author{Jonathan J. Finley$^1$}
 \email{finley@wsi.tum.de}

\affiliation{%
 $^1$Walter Schottky Institut, Department of Physics and MCQST, Technische Universität München, Am Coulombwall 4, 85748 Garching, Germany
}%
\affiliation{%
 $^2$Walter Schottky Institut, Department of Electrical and Computer Engineering and MCQST, Technische Universität München, Am Coulombwall 4 1/III, 85748 Garching, Germany
}%
\affiliation{%
 $^3$Physics Department and Solid State Institute, Technion–Israel Institute of Technology, Haifa 32000, Israel
}%
\affiliation{%
 $^4$Universit\'{e}  Grenoble  Alpes,  CNRS,  Grenoble  INP,  Institut  N\'{e}el,  38000  Grenoble,  France
}%
\affiliation{%
 $^5$Ruhr-Universität Bochum, Universitätsstraße 150, 44801 Bochum, Germany 
}%
\affiliation{%
 $^6$Technische Universität Berlin, Hardenbergstraße 36, 10623 Berlin, Germany 
}%
\affiliation{%
 $^7$Universität Paderborn, Warburger Str. 100, 33098 Paderborn, Germany  
}%

\date{\today}

\begin{abstract}
Tunnel-coupled pairs of optically active quantum dots - quantum dot molecules (QDMs) - offer the possibility to combine excellent optical properties such as strong light-matter coupling with two-spin singlet-triplet ($S-T_0$) qubits having extended coherence times. The $S-T_0$ basis formed using two spins is inherently protected against electric and magnetic field noise. However, since a single gate voltage is typically used to stabilize the charge occupancy of the dots and control the inter-dot orbital couplings, operation of the $S-T_0$ qubits under optimal conditions remains challenging. Here, we present an electric field tunable QDM that can be optically charged with one ($1h$) or two holes ($2h$) on demand. We perform a four-phase optical and electric field control sequence that facilitates the sequential preparation of the $2h$ charge state and subsequently allows flexible control of the inter-dot coupling. Charges are loaded via optical pumping and electron tunnel ionization. We achieve one- and two-hole charging efficiencies of 93.5±0.8 \% and 80.5±1.3 \%, respectively. Combining efficient charge state preparation and precise setting of inter-dot coupling allows control of few-spin qubits, as would be required for on-demand generation of two-dimensional photonic cluster states or quantum transduction between microwaves and photons. 

\end{abstract}

\maketitle


\section{\label{sec:level1}Introduction}

Long coherence times, strong light-matter coupling and tunability lie at the heart of spin-photon interfaces required for distributed quantum technologies \cite{Awschalom2018}. Semiconductor quantum dots (QDs) provide these characteristics due to their robust polarization selection rules that allow mapping between spin and optical polarization \cite{Bayer2002,Gao2012,Stockill2017}, dominant emission into the zero-phonon line at low temperatures \cite{Favero2003}, nearly Fourier-limited optical linewidths \cite{Kuhlmann2015} and integratability into devices to facilitate tunability and enhance single spin-photon coupling efficiencies \cite{Hennessy2007}. Together, these properties make QDs promising as spin-photon interfaces \cite{Lodahl2018} for the on-demand generation of 1D photonic cluster states \cite{Lindner2009,Schwartz2016} or quantum transduction between microwave and infrared photons \cite{Tsuchimoto2021}.

The growth of vertically stacked pairs of tunnel-coupled dots - quantum dot molecules (QDMs) - opens the way to form multi-spin qubits that are less susceptible to decoherence \cite{Jennings2020}. In particular, the singlet-triplet ($S-T_0$) logical qubit formed by two spins \cite{Lidar1998,Hiltunen2015, Burkard2021} occupying the hybridized orbitals of tunnel-coupled dots is expected to provide more than an order of magnitude longer coherence times ($T_2^{(*)}$) due to the existence of a sweet spot at which the $S-T_0$ qubit energy is insensitive to magnetic and electrical noise. This expectation has been confirmed for both optically active \cite{Weiss2012} and electrostatically defined QDs \cite{Bluhm2011}. QDMs have also been theoretically suggested to facilitate the deterministic (on-demand) generation of two-dimensional photonic cluster states \cite{Economou2010}, a key resource needed for measurement-based quantum computation \cite{Raussendorf2001} and memory-free quantum communication \cite{Azuma2015,Buterakos}. To facilitate these applications, QDMs must be operated in a regime where they are stably occupied by two spins, while the inter-dot tunnel coupling of s-orbital states can be freely tuned, e.g. using the voltage applied to a gate electrode ($V_G$). Thus, the charge occupancy of bottom ($S_B$) and top ($S_T$) dots in the molecule should remain in the $(S_B,S_T)\in\{(2,0), (1,1), (0,2)\}$ subspace and the tunnel coupling of the orbital states should be freely tunable without changing $n=S_B+S_T=2$. In previous experiments, all requirements have been demonstrated individually \cite{Greilich2011,Ortner2005,Scheibner2009,Krenner2005,Delley2017}. However, independent control of $n$ and tunnelling induced hybridization of orbital states has been difficult to achieve until now since $n$ is typically regulated via Coulomb blockade of carrier tunneling from a proximal doped contact \cite{Weiss2012,Kim2011} using a single gate electrode. The same gate electrode is also used to tune the orbital states into resonance, the proximity of the QDM to the doped contact, dot-dot spacing ($s$) and height ($h$) of the two dots forming the molecule must be precisely controlled during growth to allow for tunnel coupling between the s-orbital states in the QDM while remaining in the $n$=2 charge stability region.  

Here, we demonstrate all-optical, sequential and independent preparation of the $n=1$ and $n=2$ charge states of the QDM in a device geometry that leaves the gate potential free to control the orbital tunnel coupling between the two dots. Our device geometry is an n-i-Schottky diode with the QDMs embedded at the midpoint of the i-region (see Experimental Section). An AlGaAs tunneling barrier is grown $5$ nm above the QDM layer to inhibit hole tunneling escape from the QDM, while electrons can freely escape at a rate determined by $V_G$. We have previously used similar approaches to achieve selective optical charging of single QDs with electrons or holes \cite{Heiss2354,Bechtold2015}.
Consequently, using our approach spin state preparation and control is possible at precisely and arbitrarily adjustable coupling conditions by controlling the polarization and frequency of the optical charging laser relative to the discrete absorption resonances of the QDM. Moreover, we demonstrate that the optical charging process can be repeated to sequentially switch from the $n=0$ to $1$ to $2$ hole state, opening the way to access $S-T_0$ logical qubits that are insensitive to magnetic and electric field noise to first order while being fully tunable within the two-spin logical state space.\cite{Weiss2012}

\section{Results}
\subsection{Measurement Scheme}

\begin{figure*}
\includegraphics{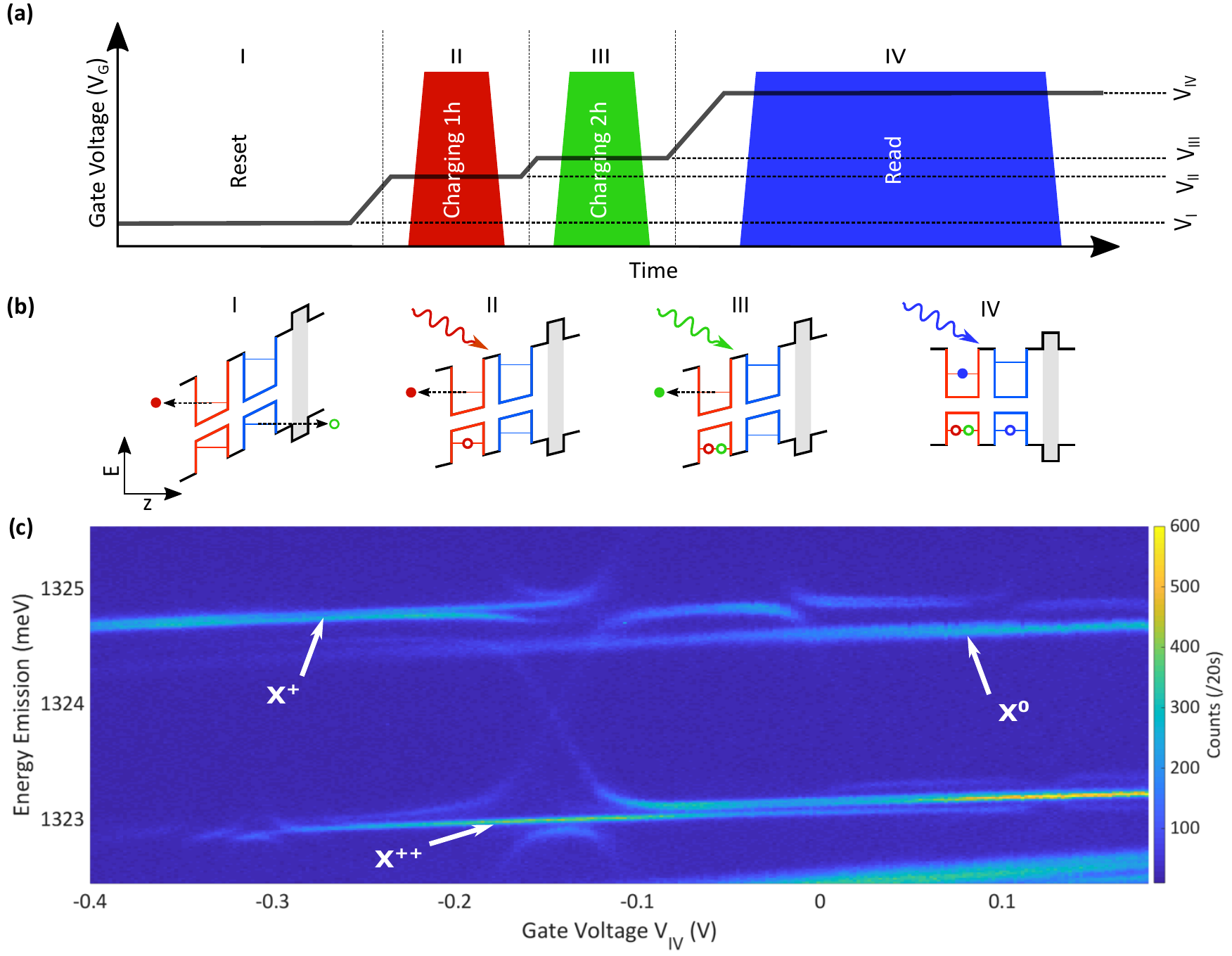}
\caption{\label{fig:wide} Optical charging of QDM. (a) Measurement scheme for charging and probing two hole states, consisting of four phases: I Reset, II Charging one hole ($1h$), III Charging two holes ($2h$), IV Readout. The black line symbolizes the gate voltage $V_G$ with voltage plateaus $V_I$ to $V_{IV}$. The colored boxes indicate laser pulses for charging and readout. (b) Band structure of QDM with double potential well (red, blue) and tunneling barrier (grey). Four sequence phases are illustrated: I - Reset: Low gate voltage leads to tunneling of electrons and holes to empty QDM, II, III - Charging one/ two holes: Resonant laser pulse creates electron hole pair. Via tunneling ionization, charges are separated and QDM loaded with one/ two holes, IV Readout: Resonant s- or p-shell excitation is applied to probe charge state. (c) Voltage dependent photoluminescence measurement showing $X^0$, $X^+$ and $X^{++}$ transitions.}
\end{figure*}

The sample investigated was grown by solid-source molecular beam epitaxy (MBE) and is an n-i-Schottky photometer with a $315$ nm thick i-region. Two layers of vertically stacked self-assembled InAs QDs were grown at the mid-point of the i-region and have a wetting layer-to-wetting layer spacing of $s=7.6$ nm. The dot height was precisely fixed at $h=2.2$ nm using the In-flush method \cite{Wasilewski1999}.  After growth of the top QD layer, a $10$ nm thick GaAs capping layer was deposited before a $20$ nm thick Al$_x$Ga$_{(1-x)}$As tunnel barrier was grown ($x=0.33$). As depicted schematically in \textbf{Figure 1b}, this barrier serves to prolong the tunneling time of the hole compared to the electron, thereby allowing selective optical charging \cite{Heiss2354}. 
Since both n-contact and the surface metallic electrode are $\geq$ 100 nm away from the QDM, tunneling induced charging from the contacts into the QDM is inhibited. Thereby, the QD molecule in our sample is largely decoupled and the optically prepared $S_T+S_B$ charge state remains unaffected by tunneling from the contacts for the bias conditions used during operation.

Our measurement scheme for charging and probing a two hole state is illustrated schematically in Figure 1a. It consists of four phases: Reset (I), charging of first hole (II), charging of second hole (III) and readout (IV). The four phases are schematically depicted in Figure 1b. During phase I, a strong reverse bias of $V_G=V_I=-4V$ is applied for $550$ ns. $V_G$ induces an axial electric field $F=(V_G-V_{BI})/d_I$ along the growth direction, where $V_{BI}$ is the built-in voltage and $d_I$ is the thickness of the intrinsic region. A strong electric field, as applied in phase I, facilitates fast tunneling escape of both electrons and holes from the molecule to initialize the QDM into the neutral state, with $n=0$ holes. In phase II of the measurement protocol, the gate voltage is tuned to $V_{II}$ in the range $\leq-0.8$ V to produce electric field conditions for which the photogenerated electron tunnels out of the molecule on timescales faster than the neutral exciton lifetime, while the hole remains stored \cite{Heiss2354}. Phase II lasts $400$ ns during which a $200$ ns laser pulse tuned resonant to the neutral exciton transition ($X^0$) in the bottom QD is gated \emph{on} using an acousto-optical modulator (AOM - red colored pulse in Figure 1a). The resonant nature of the excitation combined with the discrete electronic structure of the QDM ensures that the $n=1$ charge state is reached. Hereby, a maximum of one single electron-hole pair is generated in the QDM and, thereby, the dot remains optically active until charged by a single hole, whereupon the absorption of the trion shifts out of resonance with the driving laser field. In this way, the resonant excitation ensures that only a single hole is generated.
Once the $n=1$ hole charge state has been reached, the discrete absorption of the QDM shifts to one of the positively charged trion transitions. As such, the molecule becomes photosensitive again by tuning the driving laser frequency or switching $V_G$ to induce a DC Stark shift and re-establishing resonance with the positive trion ($X^+$). Thus, as depicted schematically in Figure 1b (III), moving from the $n=1$ to $n=2$ charge configuration involves switching the laser frequency \textit{or} electric field to a new value ($V_G=V_{III}$) whereby $X^+$ is resonantly excited. As before, a photon is absorbed whereupon the photogenerated electron tunnels out of the QDM leaving two holes in the system. Once the QDM has been optically charged with two holes, the voltage is switched to a higher level ($V_{IV}\geq-0.6$ V), for which electrons no longer tunnel out of the molecule and quantum optical experiments such as luminescence or resonance fluorescence can be performed to confirm the presence of the optically generated hole(s) in the QDM. Figure 1b (IV) depicts the readout of the charge state, denoted phase IV. Readout of the charge status $n$ in the QDM is performed by tuning a third laser field into resonance with an excited state transition of either $X^0$, $X^+$ or the doubly charged positive trion ($X^{++}$) to pump a luminescence recycling transition. Due to the quantum confined stark effect (QCSE) \cite{Fry2000}, the laser energies used to charge the QDM are significantly frequency detuned by $\Delta = -1470$ GHz ($\gg \delta \sim$10 kHz linewidth), from the readout laser allowing clean spectral filtering between charging and readout signals. As soon as two holes are prepared inside the QDM, the gate voltage is widely adjustable during the readout phase IV, e.g. to control spin-spin orbital coupling in the $2h$-molecule and explore the $S-T_0$ qubit state space. The boundaries are imposed by the voltage where electron tunneling is fast compared to the measurement time in reverse bias ($-0.6$ V) and the diode is flooded with carriers in forward bias ($0.7$ V). Within this range, any coupling condition can be addressed since the charge state is pre-set. For the measurements presented here the scheme presented in Figure 1a is continuously repeated at 420 kHz.

Depending on the charge state required we switch between $n=0$, $1$ and $2$ hole charging. This is done by either gating \textit{off} the charging laser or tuning the gate voltage such that the laser does not match the resonance condition with either the $X^0$ transition ($1h$) or the $X^0$ and $X^+$ transitions ($2h$). 
To identify readout transitions for probing zero, one, and two hole states, we recorded voltage dependent photoluminescence data. Figure 1c shows typical voltage dependent photoluminescence recorded under non-resonant excitation into the wetting-layers ($1458$ meV) as a function of the DC voltage ($V_{IV}$) applied to the Schottky contact. For this measurement the reset phase I is applied, while the charging pulses in phases II and III of the measurement scheme are gated \textit{off}. By exciting electron hole pairs in the wetting layers, charging of the QDM occurs probabilistically. Thus, the charge occupancy statistically fluctuates leading to photoluminescence signals from different charge states in the time integrated spectrum that allows simultaneous monitoring of different charge states. Crossings and avoided crossings characteristic for QDMs are observed. They arise due to the orbital hybridisation of hole states. Hybridisation takes place in both, ground and excited state, leading to charge state specific patterns \cite{Stinaff2006,Doty2008}. In this way $X^0$, $X^+$, and $X^{++}$ transitions are identified and marked in Figure 1c. These transitions link the $n$=$0$, $1$ and $2$ hole ground to excited states and are therefore used to probe the resulting charge state of the QDM after switching on the $1h$ and / or $2h$ charging laser pulses.

\subsection{One Hole Charging}

\begin{figure}
\includegraphics{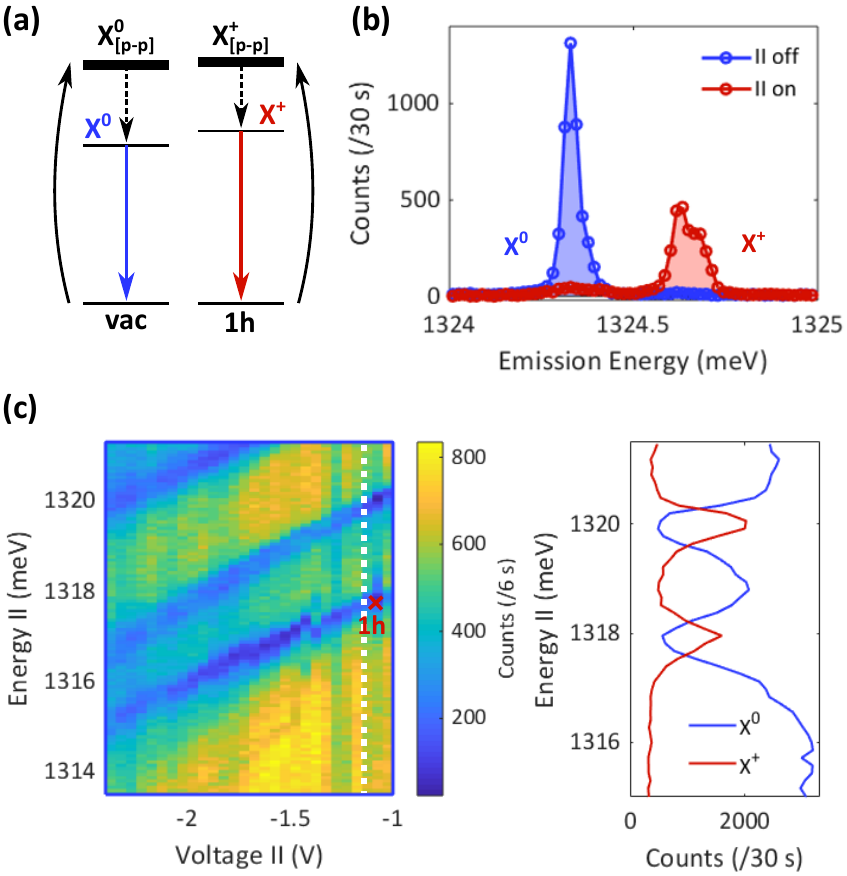}
\caption{\label{fig:epsart} Emission of $X^0$ and $X^+$ when charging one hole. (a) Excitation scheme used to identify $1h$ charging resonance. Due to overlapping p-shells, $X^0$ (blue)/ $X^+$ (red) is detected when vacuum (vac)/ $1h$ ground state is predominant. (b) Emission spectrum at $-0.2$ V under p-shell excitation with and without charging pulse of phase II applied (red/ blue). $X^0$ transition fades when charging takes places, while $X^+$ rises. (c) 2D colormap showing the $X^0$ intensity as a function of $V_{II}$ and $\hbar\omega_{1h}$. Yellow (blue) indicate high (low) count rates during the readout phase IV of the measurement. The red cross indicates the settings used in the rest of the manuscript for $1h$ charging. White dashed line marks $-1.1$ V, from where the line cut is extracted (right). Example of the anti-correlated integrated intensities of $X^0$ and $X^+$ transitions when modifying the energy of the $1h$ charging laser to identify charging resonances at phase-II of the measurement (right).}
\end{figure}

To evaluate the performance of the all-optical $1h$ charging scheme outlined above, we implemented the experimental protocol depicted in Figure 1a including only the $1h$ charging pulse (II), while omitting the $2h$ charging pulse (III). During the readout phase IV of the measurement the QDM was excited via a luminescence cycling p-p transition. In this, as well as in the following experiment, we drive transitions where the excited state electron is located in the lower quantum dot. As depicted in \textbf{Figure 2a}, the neutral exciton p-p transition ($X^0_{[p-p]}$) and the positive trion p-p transition ($X^+_{[p-p]}$) are energetically overlapping. This provides a cycling transition for both, the vacuum and the $1h$ ground state. Even if the excitation is not charge state selective, the resulting emission is. The $X^0$/ $X^+$ emission obtained when exciting the vacuum/ $1h$ ground state are separated spectrally. Therefore, we can deduce from the emission intensities on the predominant charge ground state.

The data presented in Figure 2b specifically compare measurements performed with (blue) and without (red) resonant $1h$ charging pulse applied. $V_{IV}$ was set to $-0.2$ V. If the $1h$ charging laser is blocked then the $X^0$ emission is observed, since the QDM is remains uncharged. In contrast, upon gating the charging laser $\textit{on}$ during phase II of our measurement cycle, the $X^0$ emission signal vanishes during the readout phase of the measurement. This reflects the fact that, if charging has occurred the discrete excited state of the charge neutral exciton $X^0_{[p-p]}$ is no longer optically active. On the other hand, when the $1h$ charging pulse is blocked, $X^+$ emission is not observed whereas gating $\textit{on}$ the $1h$ charging laser results in an anti-correlated increase in the intensity of $X^+$ emission at the expense of $X^0$ emission. These observations clearly indicate that selective, all optical single hole charging of the molecule has taken place.

To identify $X^0$ resonances for charging one hole, the integrated intensity of $X^0$ and $X^+$ were monitored as a function of the $1h$ charging laser frequency ($\omega_{1h}$), with the device biased at the charging voltage $V_{II}$. Figure 2c shows a 2D false-color map of the integrated emission intensity from $X^0$ for varying $V_{II}$ and $\hbar\omega_{1h}$ (left). The white dashed line marks $V_{II}=-1.1$ V, where a line cut of $X^0$ and $X^+$ emission along energies between $1315$ meV and $1321.5$ meV is recorded and presented in the rightmost panel. When the $1h$-charging laser matches the energy of a transition related to $X^0$, an electron hole pair is generated and single hole charging occurs. A transition related to $X^0$ can include s-orbital and excited state $X^0$ transitions of both dots forming the molecule, as well as indirect exciton transitions. Tunnel ionization leaves one hole in the ground state resulting in a weakening of the $X^0$ emission observed during the readout phase of the measurement. This occurs at $1318$ meV and $1320$ meV as shown in the line cut presented in Figure 2c. Concurrently, an anti-correlated increase of the $X^+$ emission is expected and observed as a fingerprint of the deterministic single hole charging. Thus, monitoring $X^0$ and $X^+$ emission intensities allows identification of $1h$ charging transitions in a voltage regime where charging takes place.

Besides adjusting the frequency of the charging laser the optimal charging voltage has to be identified. For increasingly negative $V_G$ the axial electric field becomes larger and hole tunneling times become shorter than the temporal width of the charging plateau (phase II).  A similar statement applies to the electron tunneling times becoming too long as $V_G$ becomes more positive and the axial electric field reduces. For both cases, the probability of selective charging during phase II of our measurement protocol reduces. A compromise between efficient electron tunneling and sufficiently long hole retention times is found by sweeping $\hbar\omega_{1h}$ for different $V_{II}$ (Figure 2c, left). The detected resonance energies reduce with decreasing $V_{II}$ due to the DC Stark effect. In addition, as the electron tunneling becomes faster the lifetime of the excited state decreases, broadening the linewidth of the charging resonance. By analyzing the voltage-dependent full-width half-maximum of the resonance at $V_{II}$, we estimate the electron tunneling time to be $\leq$ 2 ps \cite{Oulton2002}. 

Based on the data presented in Figure 2 and the above discussion, we select V$_{II}=-1.08$ V while the charging laser energy is fixed at $\hbar\omega_{1h}=1317.8$ meV to generate a single hole in the molecule. This optimal working point is marked by a red cross in Figure 2c and are used in the next section for sequential $2h$ charging of the QD-molecule.

\subsection{Two Hole Charging}

\begin{figure*}
\includegraphics{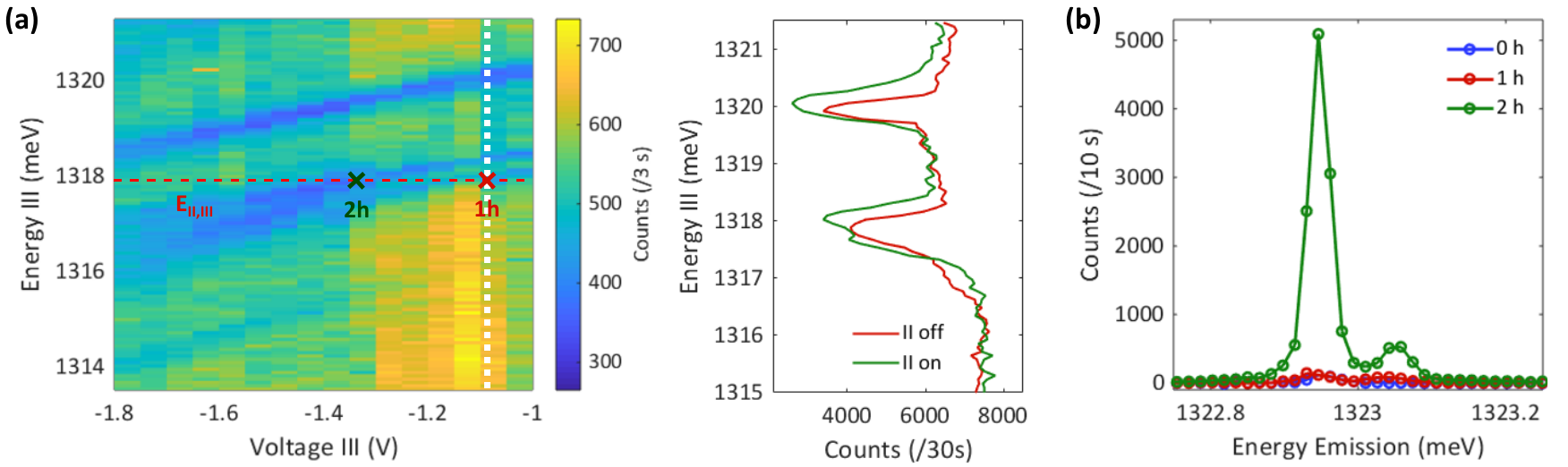}
\caption{\label{fig:wide}Two hole charging resonances. (a) $V_{III}$ and $\hbar\omega_{2h}$ dependent $X^0$ emission (left), showing shifted resonances for $2h$ charging, compared to Figure 2c. The red (green) cross indicates the voltage used for one (two) hole charging. The red dashed line illustrates the required charging energy. White dashed line marks $V_{III}=-1.08$ V at which a line cut is presented (right). Comparison of $X^0$ emission with (green curve) and without (red curve) $1h$ charging laser applied, when sweeping phase III laser energy. The charging resonances shift as the system switches between resonantly addressing the $X^0$ and $X^+$ transition for charging on and two holes, respectively. (b) Resonance fluorescence emission of the $X^{++}$ transition for charging zero / one/ two holes (blue/ red/ green).}
\end{figure*}

As discussed above in relation to Figure 1, sequential optical charging of the QDM from $n=0$ to $1$ and $2$ is achieved by switching on the charging laser during phase III of the measurement protocol. Analogously to the $1h$ charging experiment discussed in the previous section, we repeated the optimization procedure to find $V_{III}$ and $\hbar\omega_{2h}$, while using the optimal parameters $V_{II}$ and $\hbar\omega_{1h}$ found for 1h charging. The green and red curves presented in the line cut of \textbf{Figure 3a} show the integrated emission intensity of $X^0$ as a function of the laser energy in phase III of our protocol ($\hbar\omega_{2h}$), with and without the $1h$ charging laser having previously been applied in phase II. $V_{III}$ is set to match $V_{II}$, making the charging resonances of both phases comparable.  Since $X^+$ and $X^{++}$ optical transitions partly spectrally overlap, the two hole charging resonances are in a first step identified indirectly via the reduction of the $X^0$ transition intensity. Reducing the laser power of both charging pulses to $60$ \% of saturation ($240$ nW) allows to monitor the $X^0$ emission, even if charging takes place. The red curve shows resonances (e.g. at $1317.8$ meV) observed when applying only the second charging pulse. As in the data presented in Fig. 2c, $1h$ is charged by the excitation of $X^0$, leading to the observation of the same resonances. In contrast, when previously applying the first charging pulse (green curve), the second pulse can only excite the positive trion. As the $X^+$ energy is shifted compared to the $X^0$ energy, we observe a shift of resonances (e.g. at $1318$ meV). The $X^+$ resonance is then used to charge a second hole into the QDM.

Similar to situation discussed in the context of Fig. 2c, the DC Stark shift of the $X^+$ transition is observed when recording the intensity of the $X^0$ emission while varying $V_{III}$ and $\hbar\omega_{2h}$. Typical results are presented in Figure 3a. The energy and voltage of phase II are chosen as described above. A comparison with Figure 2c helps to identify voltages for which sequential $1h$ and $2h$ hole charging is achievable for the same resonant laser energy. This allows to sequentially charge two holes into the QDM with only one laser energy by modifying the gate voltage while the resonant charging laser is gated \textit{on}. Utilizing one laser only, as done in the following experiments is desirable to reduce the complexity of the charging scheme. When maintaining the $1h$ charging energy $\hbar\omega_{1h}=1317.8$ meV, the voltage of phase III has to be set to V$_{III}= -1.34$ V to enable the sequential charging of a second hole during phase III. The conditions to have sequential resonances between the charging laser and the $1h$ and $2h$ charging voltages are indicated in Figure 3a by red and green crosses, respectively. 

Up to now, the charging of a second hole has been demonstrated only indirectly via the reduction of the $X^0$ emission. In the following, the presence of two holes in the QDM is directly verified by performing resonance fluorescence on the doubly charged exciton transition $X^{++}$. In contrast to the measurements discussed above, where readout was performed via luminescence by pumping an excited state of the neutral ($X^0_{[p-p]}$) or positively charged exciton ($X^+_{[p-p]}$), here we resonantly excite and probe the s-shell doubly charged exciton $X^{++}$ transition during phase IV of our measurement. To filter out the readout laser, a cross-polarized resonance fluorescence setup was used \cite{Kuhlmann2013}. Figure 3b shows the $X^{++}$ emission upon charging zero, one, and two holes into the QDM in phases II and III of the measurement. For all three datasets the charging laser remained \textit{on} throughout phases II and III of the measurement. $V_G$ was adjusted to facilitate either $0h$, $1h$, or $2h$ charging. When charging zero and one hole(s) the signal consists mainly of unsuppressed laser. However, as soon as the QDM is loaded with two holes, the $X^{++}$ emission rises by a factor of $>18$. Besides the main emission peak at 1322.95 meV, a side peak at $1323.05$ meV is observable. This side peak is identified as an indirect $X^{++}$ transition by comparison with Figure 1c. The observed increase of the emission intensity proves that $2h$ selective optical charging of the QDM has occurred. 

Even if $1h$ and $2h$ charging is demonstrated, the experiments presented do not conclusively show that holes are sequentially added to the QDM, i.e. that the charge status is sequentially shifted from $n=0$ to $1$, to $2$ using optical pumping. To prove the sequentiality of the charging in our experiments and determine the typical photon fluxes required for charging, we continue to show that charging takes place primarily during the intended phases of the voltage sequence (II and III) and, thereby, confirm the sequential, all optical preparation of the $2h$ charge state of the QDM.

\subsection{Sequential Charging}

\begin{figure}
\includegraphics{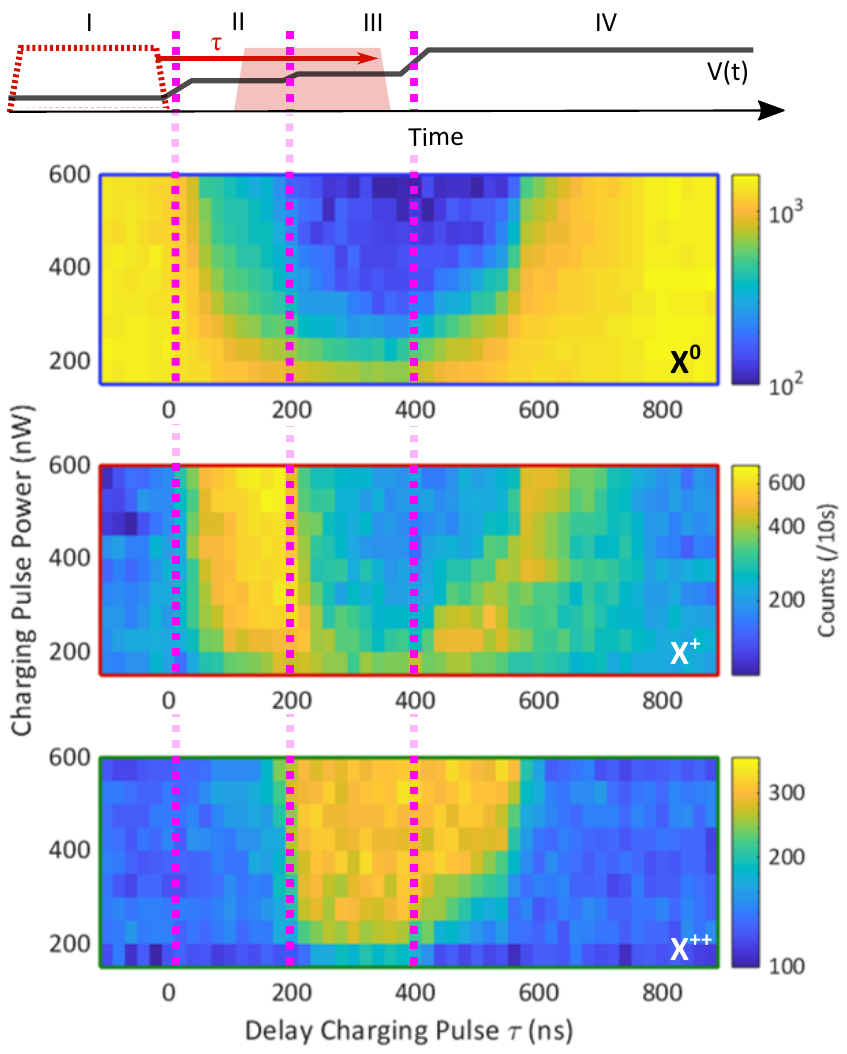}
\caption{\label{fig:epsart} Sequential two hole charging. Measurement sequence, visualizing the temporal sweep ($\tau$) of the 400 ns charging pulse over the two 200 ns charging voltage plateaus. At $\tau$ = 0 ns, the charging laser enters the first charging plateau. The readout is performed via p-shell excitation while detecting the spectrally detuned s-shell emissions of $X^0$, $X^+$ and $X^{++}$. Below: Integrated emission of $X^0$, $X^+$ and $X^{++}$ (top to bottom) is logarithmically plotted for varying charging pulse delay and charging power. Magenta dashed lines indicate the edges of the $1h$ and $2h$ charging plateaus.}
\end{figure}

To confirm sequentiality of the charging process, we performed a measurement using a single charging pulse having a temporal width of $400$ ns.  This charging pulse was swept through the two charging plateaus of phase II and III, each of which is $200$ ns long. This procedure is illustrated schematically in \textbf{Figure 4}, showing the measurement sequence and the temporal advance of the charging laser pulse $\tau$. At $\tau<0$, the charging laser overlaps solely with phase I of the measurement sequence and is, therefore, ineffective in charging the QDM. The charging pulse is then shifted through phases II ($1h$ charging), III ($2h$-charging) of the measurement protocol until it overlaps solely with phase IV (readout). By advancing the time when the charging pulse is applied, its temporal overlap with the two charge plateaus changes. Figure 4 shows the emission of $X^0$, $X^+$ and $X^{++}$ transitions as a function of the temporal advance $\tau$ and the applied charging power. The charging energy and the voltage plateaus of phase II and III are set as previously defined for two hole charging. In addition, the second charging laser was turned \textit{off} for this experiment. Readout was performed via luminescence driven by pumping p-shell transitions at $-0.2$ V to detect $X^0$ (Figure 4 - upper panel), $X^+$ (middle panel) and $X^{++}$ (lower panel) emission simultaneously. 

Complete embedding of the charging pulse in phase I ($\tau$ \textless\ 0 ns) does not result in charging. $X^0$ therefore dominates the $X^+$ and $X^{++}$ emission. At $\tau$ = 0 ns, the laser pulse enters the $1h$ charging plateau of phase II.  As a result, the emission intensity of $X^+$ rises as charging of the QDM with a single hole takes place. At the same time, the emission intensity of $X^0$ reduces while the $X^{++}$ remains close to the background level indicative of the charge state of the QDM being enhanced to $n=1$. For low power, this effect occurs at longer values of $\tau$, which corresponds to a larger overlap between laser pulse and charging plateau. This reflects the fact that the charging efficiency is reduced - charging is probabilistic due to the tunneling process and weaker photon fluxes require longer times to establish the $n=1$ charge state. The two hole charging plateau is reached at $\tau=200$ ns. While the $X^0$ emission remains at a low level, the intensity of $X^{++}$ progressively rises with $\tau$ and the intensity of $X^+$ simultaneously reduces in an anti-correlated manner. This key observation shows that the number of charged holes sequentially increased from $n=1$ to $2$. The second hole is accordingly mainly charged during phase III due to the finite orbital degeneracy of the states excited. Up to $\tau$ = $600$ ns $2h$ charging is performed, as the laser pulse overlaps with both charging plateaus. After $\tau$ = $600$ ns the pulse leaves the first charging plateau, whereupon $n=1$ charging has not occurred anymore. As a result, the $X^+$ transition cannot be excited during phase III, and the charge status of the QDM remains close to $n=0$.  Consequently, the emission from $X^0$ reappears while $X^{++}$ progressively decreases. We attribute the small increase of $X^+$ emission with $\tau$ to the nonzero probability of charging one hole during phase III. As the resonances broaden with decreasing voltage and $X^0$ and $X^+$ charging transitions spectrally overlap, the selectivity of the charging process is reduced. This makes generation of unwanted charge states more likely.

\section{Discussion and Summary}
To quantify the efficiency of the proposed charging scheme we calculated P(n$|$m), the conditional probability for charging $n$ holes, given $m$ charging pulses are applied. The reset probability is extremely close to unity (P(0$|$0) = 1), and considering only states with n $\leq$ 2 to reflect the degeneracy of a single QD orbital, we obtain $1h$/ $2h$ charging probabilities of $\geq$ 93.5±0.8 \%/ $\geq$ 80.5±1.3 \%, respectively (see section 4.2). These results show that the sequential, all-optical preparation of desired number of charges is a robust and reliable process. Certainly, the optical preparation of charges is more complex compared to  QDM charging via tunneling from a diode contact. Additional voltage plateaus and laser pulses are required. Furthermore, the repetition rate of the sequence is currently limited to 740 kHz due to the duration of reset and charging pulses. However, this limitation is mainly caused by the RC time constant of the diode. By using micron scale photodiodes having low RC time constants \cite{Pedersen2020}, this can be increased to up to 500 MHz. Moreover, it is superior to optimize protocols for increasing the readout phase IV duty cycle. This can be done by performing several readout cycles after having previously prepared the charge state. The limit of such approaches will be determined by the extent to which measurement back action influences the charge status of the QDM, e.g. via Auger auto-ionization \cite{Lobl2020}.
Furthermore, the demonstrated scheme does not pose any restrictions on the readout voltage to obtain certain charge states. The $X^0$, $X^+$ and $X^{++}$ intensities shown in Figure 4 are acquired at the same gate voltage, showing its independence.

In summary, we have proposed and demonstrated a four-phase measurement sequence that allows independent control of charge status and inter-dot coupling of a single, electrically tunable QDM. We demonstrated one and two hole charging via optical excitation and tunnel ionization. The addressability of the generated charge state was thereafter not affected by the gate voltage, facilitating electrically-tunable spin-spin interactions induced using the exchange coupling between the two spins \cite{Petta2005}. By combining efficient charging and separate gate voltage control the proposed method of optical charging is therefore suitable to replace and outperform previous charging approaches with respect to controllability and flexibility. Independent charge state preparation paired with the ability to manipulate the inter-dot coupling paves the way for protocols requiring simultaneous spin and coupling control, as for example needed for 2D cluster state generation. \cite{Economou2010}

\section{Experimental Section}
\subsection{Sample Structure}
The QDM investigated was grown by solid-source molecular beam epitaxy. It consisted of two laterally stacked InAs quantum dots embedded in a GaAs matrix. The inter-dot coupling strength is determined by the wetting layer to wetting layer separation of $7.6$ nm. The individual dot height is fixed to $2.2$ nm via an In-flush technique. Thereby, the energies of the two QDs are adjusted. The upper QD is designed to have a higher confinement energy compared to the lower QD to facilitate electric field induced tunnel coupling of orbital states in the valence band \cite{Bracker2006}. A $20$ nm thick Al$_x$Ga$_{(1-x)}$As tunnel barrier ($x=0.33$) was grown $10$ nm above the QDM to prolong hole tunnelling times and enable tunnel ionization charge state preparation. Furthermore, the molecule is embedded into an n-i-Schottky diode to apply electric fields along the growth direction of the sample. The n-doped region and the Schottky surface metallic electrode are used as contacts. Both diode contacts are more than 100 nm away from the molecule to prevent uncontrolled charge tunnelling into the QDM and, non-optical or selective modification of the charge status. All measurements are performed at 10 K. For preparation and readout of the charge state tunable diode lasers are used.

\subsection{Fidelity Calculation}
To estimate a lower boundary of the one and two hole charging fidelities, we assume that the reset phase I is perfect: P(0$|$0) = 1. This expectation is likely to be achived in our experiment since the discharging electric field is high and the duration of the initialization phase of the measurement is sufficiently long.  Here, P($n$$|$$m$) is the probability of charging $n$ holes, given $m$ charging pulses are applied. This assumption can be justified, as duration and voltage of phase I can be chosen to make the presence of charges negligible. Furthermore, only charge states with n $\leq$ 2 are taken into account due to the orbital structure of the charged QDM.

By monitoring the $X^0$ emission with and without applying one charging pulse, we can calculate the probability of charging a non-zero number of holes P($\neg$ 0$|$1) = $\frac{N^{0}_{0}-N^{0}_{1}}{N^{0}_{0}}$. Where $N^{a}_{b}$ is the number of counts observed for transition $X^a$, with $a \in \{0,+,++\}$. $b \in \{0,\ 1,\ 2\}$ on the other hand indicates the number of charge pulses applied. 
The probability for selectively charging one hole, given that we want to charge a single hole, can be written as P(1$|$1) = P($\neg$ 0$|$1) - P(2$|$1).

The probability for charging two holes when only one is intended can be written as: P(2$|$1) $\leq \frac{N^{++}_{1}}{N^{++}_{2}}$. We normalize the counts obtained monitoring the $X^{++}$ transition when planned to charge one hole by the counts of the $X^{++}$ transition when planned to charge two holes. This indicates in which proportion of the overall cases we charge the dot with two holes if we only want to have one. In the case of a perfect two hole charging procedure, this inequality changes to an equation.

The fidelity of charging one hole can now be written as:
\begin{equation}P(1|1) \geq \frac{N^{0}_{0}-N^{0}_{1}}{N^{0}_{0}} - \frac{N^{++}_{1}}{N^{++}_{2}}\end{equation}

The two hole charging fidelity can be calculated by estimating its counter events: P(2$|$2) = 1 - P(1$|$2) - P(0$|$2). 
The probability of charging zero holes for this case is estimated via the counter event of charging zero holes, given a charging pulse is applied: P(0$|$ 2) $\leq$ 1 - P($\neg$\ 0$|$1).
The probability for loading one hole given two holes are intended, we follow the same argumentation as for calculating P(2$|$1). However, we monitor the $X^+$ transition and divide the counts of the two hole by the one hole charging case: P(1$|$2) $\leq \frac{N^{+}_{2}}{N^{+}_{1}}$.

This allows us to write the two hole charging fidelity as:
\begin{equation}P(2|2)\geq\frac{N^{0}_{0}-N^{0}_{1}}{N^{0}_{0}}-\frac{N^{+}_{2}}{N^{+}_{1}}\end{equation}
\\
\section{\label{sec:level1}Acknowledgements}
We gratefully acknowledge financial support from the German Federal Ministry of Education and Research (BMBF) via Q.Link.X (16KIS0874), QR.X (16KISQ027), the European Union’s Horizon 2020 research and innovation program under Grant Agreement 862035 (QLUSTER) and the Deutsche Forschungsgemeinschaft (DFG, German Research Foundation) via SQAM (FI947-5-1), DIP (FI947-6-1), and the Excellence Cluster MCQST (EXC-2111, 390814868). FB gratefully acknowledges the Exploring Quantum Matter (ExQM) programme funded by the State of Bavaria. F.S., K.B., and K.M. gratefully acknowledges the BMBF for financial support via project MOQUA (13N14846). 
\\

Key words: Quantum dot molecule, charge state control, inter-dot coupling, charge storage, optical charging


\bibliographystyle{MSP}
\bibliography{TwoHoleChargingRef}

\end{document}